\documentclass[aps,prf,reprint,superscriptaddress,onecolumn]{revtex4-2}
\usepackage[usenames,dvipsnames]{xcolor}
\usepackage{lipsum}
\usepackage{graphicx}
\usepackage{array}
\usepackage{epstopdf}
\usepackage[justification=justified]{caption}
\usepackage{amsmath}
\usepackage[arrowdel]{physics}
\usepackage{multirow}
\usepackage{amsmath}
\usepackage{amsfonts}
\usepackage{amssymb}
\usepackage{mathrsfs}
\usepackage{mathtools}
\usepackage{tikz}
\usepackage{pgfplots}
\usepackage{csquotes}

\definecolor{color1}{RGB}{0,113,188}
\definecolor{color2}{RGB}{216,82,24}
\definecolor{color3}{RGB}{236,176,31}

\newsavebox\mycircle
\savebox\mycircle{                                                                   
  \begin{tikzpicture}                                                           
    \tikz\draw[line width=0.3mm, color=color1] (0.05,0.05)
circle (0.5ex); 
  \end{tikzpicture}                                                             
} 

\newsavebox\mycircleblack
\savebox\mycircleblack{                                                                   
  \begin{tikzpicture}                                                           
    \tikz\draw[line width=0.3mm, color=black] (0.05,0.05)
circle (0.5ex); 
  \end{tikzpicture}                                                             
} 

\newsavebox\myline
\savebox\myline{                                                                   
  \begin{tikzpicture}                                                           
    \tikz\draw[line width=0.3mm, color=color2] (0,-0.05)--(0.5,-0.05);
  \end{tikzpicture}                                                             
} 

\newsavebox\mylineblue
\savebox\mylineblue{                                                                   
  \begin{tikzpicture}                                                           
    \tikz\draw[line width=0.3mm, color=color1] (0,-0.05)--(0.5,-0.05);
  \end{tikzpicture}                                                             
} 

\newsavebox\mylineblack
\savebox\mylineblack{                                                                   
  \begin{tikzpicture}                                                           
    \tikz\draw[line width=0.3mm, color=black] (0,-0.05)--(0.5,-0.05);
  \end{tikzpicture}                                                             
}

\newsavebox\mylinedashed
\savebox\mylinedashed{                                                                   
  \begin{tikzpicture}                                                           
    \tikz\draw[dashed ,step=0.2, color=black] (0,-0.05)--(0.5,-0.05);
  \end{tikzpicture}                                                             
}

\newsavebox\mylinedashedblack
\savebox\mylinedashedblack{                                                                   
  \begin{tikzpicture}                                                           
    \tikz\draw[dotted ,step=0.2, color=black] (0,-0.05)--(0.5,-0.05);
  \end{tikzpicture}                                                             
}

\newsavebox\mylinedashedblue
\savebox\mylinedashedblue{                                                                   
  \begin{tikzpicture}                                                           
    \tikz\draw[dashed ,step=0.2, color=color1] (0,-0.05)--(0.5,-0.05);
  \end{tikzpicture}                                                             
}

\newsavebox\mylinedashedred
\savebox\mylinedashedred{                                                                   
  \begin{tikzpicture}                                                           
    \tikz\draw[dashed ,step=0.2, color=color2] (0,-0.05)--(0.5,-0.05);
  \end{tikzpicture}                                                             
}

\newsavebox\myboxblue                                                           
\savebox\myboxblue{                                                                   
  \begin{tikzpicture}                                                           
    \draw [line width=0.3mm, color=color1] (0.1,0.1) rectangle(0.3,0.3);                                       
  \end{tikzpicture}                                                             
}

\newsavebox\myboxred
\savebox\myboxred{                                                                   
  \begin{tikzpicture}                                                           
    \draw [line width=0.3mm, color=color2] (0.1,0.1) rectangle(0.3,0.3);                                       
  \end{tikzpicture}                                                             
}

\newsavebox\myboxblack                                                           
\savebox\myboxblack{                                                                   
  \begin{tikzpicture}                                                           
    \draw [fill=black] (0.1,0.1) rectangle(0.3,0.3);                                       
  \end{tikzpicture}                                                             
}  

\newsavebox\myboxgray                                                           
\savebox\myboxgray{                                                                   
  \begin{tikzpicture}                                                           
    \draw [fill=gray] (0.1,0.1) rectangle(0.3,0.3);                                       
  \end{tikzpicture}                                                             
}   

\newsavebox\myboxwhite
\savebox\myboxwhite{                                                                   
  \begin{tikzpicture}                                                           
    \draw [fill=white] (0.1,0.1) rectangle(0.3,0.3);                                       
  \end{tikzpicture}                                                             
}

\begin{document}

\author{Igor A. Maia}
\affiliation{Divisão de Engenharia Aeroespacial, Instituto Tecnológico de Aeronáutica, São José dos Campos, 12228-900, Brazil}
\author{Maxime Fiore}
\affiliation{Department of Aerodynamics, Energetics and Propulsion, ISAE-SUPAERO, Université de Toulouse, 31400, Toulouse, France}
\author{Romain Gojon}
\affiliation{Department of Aerodynamics, Energetics and Propulsion, ISAE-SUPAERO, Université de Toulouse, 31400, Toulouse, France}

\title{Tones and upstream-travelling waves in ideally-expanded round impinging jets}

\begin{abstract}

%\lipsum[1-1]

We study the generation of tones by ideally-expanded round jets impinging on a flat plate. Data from large-eddy simulations performed for different nozzle-to-plate distances is explored, and we consider closure of the aeroacoustic feedback loop responsible for the tones by guided jet modes. Allowable frequency ranges for resonance, underpinned by the existence of modes with upstream-directed group velocities, are computed using two different models: a cylindrical vortex-sheet model; and a locally-parallel stability model which considers a finite-thickness velocity profile. It is shown that inclusion of a finite-thickness velocity profile consistent with the mean flow in the vicinity of the plate improves the agreement between observed tones and model predictions. The frequency of the largest tones found in the data are found to fall within, or very close to, the frequency limits of the finite-thickness model, correcting discrepancies observed with the vortex-sheet model. The same trend is observed in comparisons with experimental and numerical data gathered from the literature. Pressure eigenfunctions of the stability model are in good agreement with upstream-travelling disturbances educed from the data at the tone frequencies. This provides further evidence for the involvement of guided jet modes in the resonance mechanism.

\end{abstract}

\maketitle

	\section{Introduction}
	\label{sec:intro}
	
It is known that high-subsonic and supersonic jets impinging on a wall are characterised by intense acoustic tones. Following the seminal work of \citet{Powell1953}, the properties of such tones have been widely studied experimentally \cite{henderson1966experiments, neuwerth1974acoustic, wagner1971sound, powell1988sound, krothapalli_rajkuperan_alvi_lourenco_1999, henderson_bridges_wernet_2005, alvi2008experimental, kumar2009control, risborg2009high, venkatakrishnan2011density, mitchell2012visualization} and, more recently, numerically \cite{gojon_bogey_marsden_2016, bogey_gojon_2017, gojon_bogey_aiaa2018, varé_bogey_2022, vare_bogey_aiaa2023} for different Mach numbers and nozzle-to-plate distances. For subsonic and ideally-expanded supersonic impinging jets, the tones are generated by an aeroacoustic feedback mechanism, which involves hydrodynamic disturbances convected downstream that impinge on the plate and excite upstream-travelling waves. The latter are reflected by the nozzle, exciting the downstream-travelling disturbances and closing the loop. For non-ideally expanded supersonic impinging jet, they seem to be associated in some cases to the same feedback loop \cite{jaunet_aiaa2019}, and in other cases to a feedback loop happening only when a Mach disk forms just upstream from the plate, leading to "silence zones" with no tonal noise observed for some nozzle-to-plate distances \cite{henderson1993experiments,kuo1996oscillations,gojon2017flow}.

There is a relatively well-established consensus that the downstream-travelling hydrodynamic disturbances are underpinned by the Kelvin-Helmholtz instability \cite{edgington2019aeroacoustic}. The nature of the upstream-travelling waves, on the other hand, has been subject of debate. In the models of \citet{ho_nosseir_1981} and \citet{powell1988sound}, the loop is closed by free-stream acoustic waves (propagating outside of the jet column), which gives rise to simple models to predict the tone frequencies, based on hydrodynamic and acoustic phase speeds, and the nozzle-to-plate distances. Later, \citet{tam_ahuja_1990} proposed an alternative model, in which the feedback is provided by upstream-propagating neutral waves that have a spatial support \textit{inside} the jet. Unlike free-stream acoustic waves, they are dispersive and travel with a phase velocity that is slightly below the ambient speed of sound. Their existence was first postulated by \citet{tam_hu_1989}, who showed that their spatial characteristics could be obtained from a cylindrical vortex-sheet model. \citet{tam_norum_1992} then extended this model to planar jets. \citet{gojon_bogey_marsden_2016} and \citet{Bogey2017} used these models to investigate the origin of tones observed in large-eddy simulations (LES) of ideally-expanded planar and round jets, respectively, impinging on flat plates. The vortex sheet models provided a fairly good prediction for the frequency and azimuthal/spanwise organisation (axisymmetric or helical) of a number of tones. However, several tones lied outside of the allowable frequency range of existence for those modes predicted by the vortex sheet, and their origin remained unexplained.

Upstream-travelling waves supported by the jet are now known to participate in other resonant phenomena. For instance, they are involved in the resonance observed in the potential core of subsonic jets \cite{towne_cavalieri_jordan_colonius_schmidt_jaunet_brès_2017} and in the tonal dynamics of jet-flap interaction \cite{jordan_jaunet_towne_cavalieri_colonius_schmidt_agarwal_2018} noise. They are also key elements in screech resonance, as demonstrated numerically for round and planar screeching jets by \citet{gojon2018oscillation} and  \citet{gojon2019antisymmetric}, respectively; and experimentally by \citet{edgington-mitchell_jaunet_jordan_towne_soria_honnery_2018} and \citet{mancinelli2019screech} in round jets. Furthermore, \citet{mancinelli2019screech} showed that a prediction model that considers the upstream-travelling jet modes as a closure mechanism provides much better agreement with experimental data compared to that of free-stream acoustic waves, over a broad range of Mach numbers.

%A similar experimental campaign was conducted by \citet{jaunet_aiaa2019} in underexpanded round impinging jets. They compared their experimental observations to tone prediction models based on free-stream and upstream jet waves; their results provide evidence that the closure mechanism for impinging jets is also more consistent with the dynamics of the latter. This confirmed the observations of \citet{bogey_gojon_2017} for ideally-expanded jets. -> ROMAIN : j'en parle plus haut du cas non idéalement détendu, car c'est plus complexe que ce que Vincent dit dans son papier.

 Some of the difficulty in making accurate predictions comes from the lack of knowledge about the reflection coefficients of downstream and upstream travelling waves at the plate and the nozzle, as pointed out by the authors, which requires the addition of a free parameter in the model. Another source of inaccuracy in the model may come from the assumption of an infinitely thin shear layer. Despite being able to capture the general trends, vortex-sheet-based prediction models do not produce a perfect collapse with numerically-observed tones in the frequency-Mach number plane \cite{gojon_bogey_marsden_2016, Bogey2017}. It was recently shown by \citet{mancinelli_jaunet_jordan_towne_2021} that the inclusion of a finite thickness leads to significant improvements in screech-frequency predictions. They also showed that a finite-thickness model is required in order to capture temperature effects on screech generation. In light of this, and given the similarities in the dynamics of resonance loops of screeching and impinging jets, in this work we consider a finite-thickness models, based on local stability analysis, to identify allowable frequency ranges for resonance in impinging jets. We revisit the LES of ideally-expanded round impinging jets by \citet{Bogey2017} and \citet{gojon_bogey_aiaa2018}, for which vortex-sheet-based resonance predictions were recently made \cite{ferreira_aiaa2023}. These predictions revealed mismatches regarding the frequencies and spatial structures of the upstream-travelling waves computed from the model and those educed from the LES for some of the tones. We show that considering the shear layer profile in the vicinity of the plate, consistent with the velocity field of the LES, extends the eligible frequency ranges for resonance and corrects those mismatches to a great extent. 

The remainder of the paper is organised as follows: Sec.~\ref{sec:database} introduces the LES database and provides details of the numerical methods used; Sec.~\ref{sec:models} presents the vortex-sheet and finite thickness models; Sec.~\ref{sec:results} presents the comparison between LES and model predictions and is followed by concluding remarks in Sec.~\ref{sec:conclusions}.

\section{Database \label{sec:database}}

\begin{figure}
    \centering
     \def\svgwidth{0.7\textwidth} 
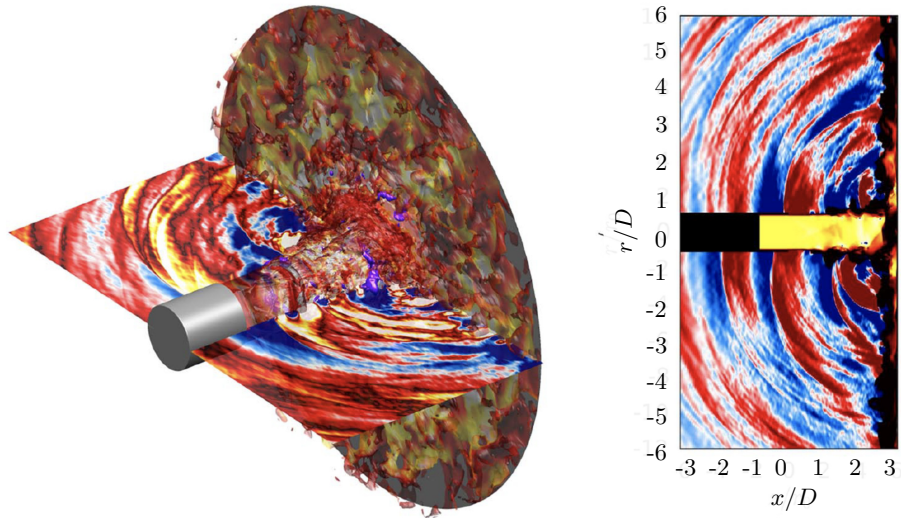
    \caption{Instantaneous fields for the Jetideal3D case. Left: isosurface of density at 1.3 kg.m$^{-3}$, colored by the Mach number showing the jet column and impingement, pressure fluctuation fields in the plane $\theta = 0^\circ$ and close to the flat plate (x constant plane). Right: contours of density in the jet and pressure fluctuations outside of the jet.The color scale ranges from 1 to 2 kg.m$^{-3}$ for the density and from -5000 to 5000 Pa for the fluctuating pressure}
    \label{visu_ideal_jet_round}
\end{figure}

We consider LES databases of ideally expanded round jets that originate from a straight pipe nozzle of diameter $D=2 R=0.002$~m, whose lip is $0.05D$ thick, in an ambient medium at temperature $T_{0}=293$~K and pressure $p_{0}=10^{5}$~Pa. 
Four nozzle-to-plate distances were simulated: $L=3D, 4D, 5D, 6D$. These cases will be referred respectively as Jetideal3D, Jetideal4D, Jetideal5D, and Jetideal6D. At the nozzle exit, the jets are ideally expanded, and have a Mach number of $M_{j}=U_{j}/a_{j}=1.5$, and a Reynolds number of $Re_{j}=U_{j} D/\nu=6 \times 10^{4}$, where $U_{j}$~=~427.3~m.s$^{-1}$ is the jet exit velocity, $a_{j}$ the speed of sound at the jet exit, and $\nu$ the kinematic viscosity. The total to ambient temperature ratio for all cases is $T_0/T_{\infty}=1$, yielding $T_j/T_{\infty}=0.69$ at this Mach number. Figure ~\ref{visu_ideal_jet_round} illustrates a snapshot of density and pressure fields for the Jetideal3D case.

%\begin{figure}
%    \centering
%\includegraphics[trim=0cm 0cm 0cm 0cm, clip=true,width=0.7\linewidth]{Figures/visu_ideal_jet_round1.pdf}
%    \caption{Instantaneous fields for the Jetideal3D case. Left: isosurface of density at 1.3 kg.m$^{-3}$, colored by the Mach number showing the jet column and impingement, pressure fluctuation fields in the plane $\theta = 0^\circ$ and close to the flat plate (x constant plane). Right: contours of density in the jet and pressure fluctuations outside of the jet.The color scale ranges from 1 to 2 kg.m$^{-3}$ for the density and from -5000 to 5000 Pa for the fluctuating pressure}
%    \label{visu_ideal_jet_round}
%\end{figure}

\begin{figure}
\centering
\def\svgwidth{0.5\textwidth}
\input{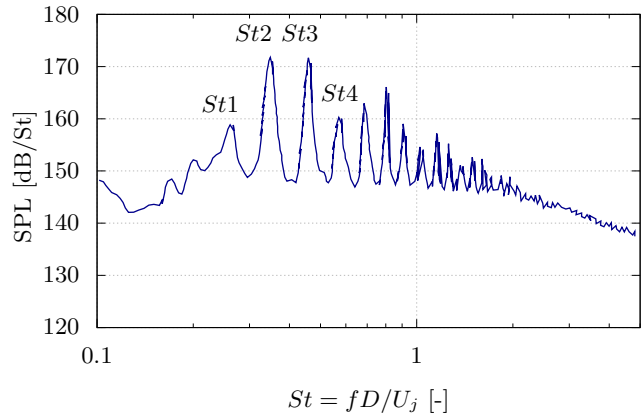}
\caption{SPL at $x/D=0$ and $r/D=1$ as function of the Strouhal number $St=fD/U_{j}$ for the Jetideal3D case. See Tab.~\ref{tab:st} for the corresponding frequencies of the tones $St_1$ to $St_4$}
\label{SPL_3dj}
\end{figure}

The unsteady compressible Navier–Stokes equations are solved in cylindrical coordinates, ($r, \theta, x$) using an explicit six-stage Runge–Kutta algorithm for time integration and low-dissipation and low-dispersion explicit 11-point finite differences for spatial derivation~\cite{Bogey2004,Berland2007}. At the end of each time step, a high-order filtering is applied to the flow variables to remove grid-to-grid oscillations and to dissipate subgrid-scale turbulent energy~\cite{Bogey2006, Bogey2009,Fauconnier2013}. The cylindrical meshes contain between 202 and 240 million points. The minimal axial mesh spacing, equal to $\Delta x/D=0.00375$, is located near the nozzle lip and the flat plate, and the maximal axial mesh spacing, equal to $\Delta x/D=0.015$, is located between the nozzle and the plate. The minimal radial spacing is equal to $\Delta r/D=0.00375$ at $r= D/2$, and the maximal radial spacing, excluding the sponge zone, is $\Delta r/D = 0.03$ for $2.5 D \le r \le 7.5D$. The maximum mesh spacing of $\Delta r/D=0.03$ allows acoustic waves with Strouhal numbers up to $St=fD/U_{j}=5.3$ to be well propagated in the computational domains, where $f$ is the frequency. Numerical details can be found in Bogey and Gojon~\cite{Bogey2017}, and the flow field is discussed in greater detail in \citet{gojon_bogey_aiaa2018}.
Figure~\ref{SPL_3dj} shows a sound pressure level (SPL) corresponding to SPL~=~$20 \log_{10}(p/p_{\mathrm{ref}})$ with $p_{\mathrm{ref}}=2 \times 10^{-5}$~Pa computed at $x/D=0$ and $r/D=1$ for the Jetideal3D case, as a function of the Strouhal number. The spectrum is characterised by several tones that rise, at some frequencies, orders of magnitude above the broadband noise level. Pressure spectra computed in the near acoustic fields of the other simulated cases (not shown) display similar features. The Strouhal numbers of the dominant tones in each database are displayed in Tab.~\ref{tab:st}. The azimuthal organisation of the pressure field at the tone frequency (either axisymmetric or helical), verified through an azimuthal Fourier decomposition of the signal, is also indicated.

%\begin{figure}
%\centering
%\includegraphics[trim=0cm 0cm 0cm 0cm, clip=true,width=0.5\linewidth]{Figures/SPL_3dj.png}
%\caption{Sound pressure levels (SPLs) at \bm{$r=2r_{0}$} and \bm{$x=0$} as functions of the Strouhal number \bm{$St=fD/u_{j}$}}
%\label{SPL_3dj}
%\end{figure}
%

\begin{table}[hbt!]
\caption{\label{tab:table1} Strouhal numbers of the main tones emerging in the acoustic spectra. (A) and (H) indicate axisymmetric and helical oscillation modes, respectively. The dominant tone for each case is highlighted in bold.}
\hspace*{-0.5cm}
\centering
\begin{tabular}{ccccc}
\hline
 & St1 & St2 & St3 & St4 
\\ \hline
Jetideal3D & 0.26 (H) & 0.345 (H) & \textbf{0.455} (A) & 0.57 (H) \\
Jetideal4D & 0.205 (A) & 0.29 (H) & 0.365 (A) & \textbf{0.445} (A) \\
Jetideal5D & 0.165 (A) & 0.29 (H) & 0.375 (A) & \textbf{0.44} (A) \\
Jetideal6D & 0.175 (A) & 0.255 (H) & 0.305 (H) & \textbf{0.38} (A) \\

\end{tabular}
\label{tab:st}
\end{table}

\section{Models \label{sec:models}}

In this section, we present prediction models for possible frequency ranges of resonance. We assume that the feedback loop involves a downstream-travelling Kelvin-Helmholtz wave and an upstream-travelling guided jet mode. Here we focus on the characterisation of the latter, as it is the one delimiting the allowable frequency range for resonant loop to occur. It appears in families of waves organised by azimuthal wavenumber, $m$ and radial order, $n$, depending of the number of anti-nodes in their radial structure. Resonance is only possible in limited frequency ranges where such modes are propagative (neutral). For supersonic jets, this region is delimited by the saddle and branch points of their dispersion relation \cite{Tam1990}. Two models are used to compute the dispersion relations of those modes: an infinitely thin (vortex-sheet) model used in a previous study of these impinging jets~\cite{gojon2018oscillation}; and a spatial stability model based on the Euler equations. 

We emphasise that our goal is not to provide a model that predicts the tonal frequencies \textit{exactly}. This would require a wavenumber-matching criterion that involves knowledge of the reflection coefficients at the nozzle and plates \cite{mancinelli2019screech, mancinelli_jaunet_jordan_towne_2021, jaunet_aiaa2019}, which are generally unknown. Moreover, designing an accurate tone-prediction model presumes tracking the tones in the frequency-Mach number plane (as done in the above-mentioned screech studies), which requires data at a large number of jet operating conditions. These two aspects are beyond the scope of this work; here we focus on computing \textit{eligible} frequency ranges for resonance. Particularly, we wish to assess whether inclusion of a finite shear-layer thickness can provide a possible explanation for some tones observed in the data that were found to be inconsistent with a vortex-sheet model \cite{Bogey2017,ferreira_aiaa2023}.

\subsection{Cylindrical vortex-sheet model}
%\begin{figure}
%    \centering
%     \def\svgwidth{0.5\textwidth}
%\input{Figures/scheme_vs_model1.eps_tex}
%    \caption{Two-dimensional supersonic jet bounded by vortex sheets}
%    \label{fig:scheme_vs_model}
%\end{figure}
\begin{figure}
    \centering
     \def\svgwidth{0.7\textwidth} 
\input{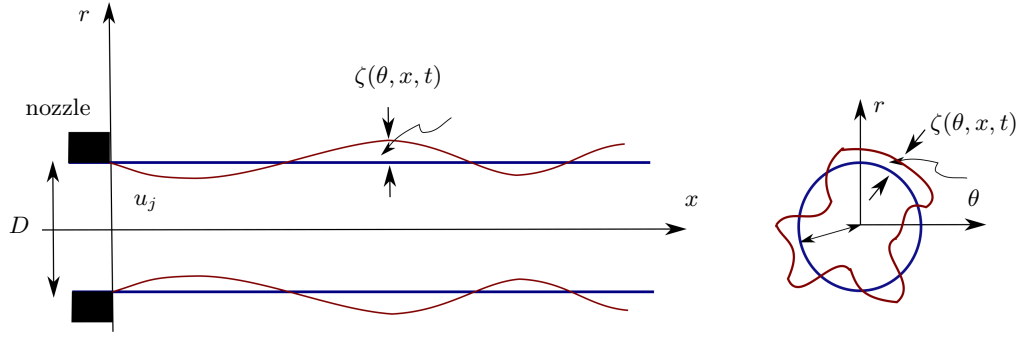}
    \caption{Perturbed motion of a round jet bounded by a vortex sheet.}
    \label{fig:shear_layer_model1}
\end{figure}

The dispersion relations using a vortex sheet model for a round jet introduced in this Sec. follow the initial derivation provided in Tam (1971)~\cite{Tam1971} and also used later in Tam and Hu (1989)~\cite{Tam1989}. The model considers the streamwise evolution of a small-amplitude disturbance wave in an infinitely thin shear layer (see Fig.~\ref{fig:shear_layer_model1}). The disturbance is assumed to be initiated by a localized excitation of frequency $\omega$, giving rise to an instability wave. This excitation can be produced by external forcing when acoustic waves generated by turbulent structures impinging on the flat plate can travel upstream up to the end of the subsonic region in the stagnation region, for instance. The static pressure is assumed to be constant throughout the flow. The instability wave and its acoustic field will be assumed to satisfy the linearised inviscid, compressible equations of motion. The three-dimensional vortex sheet model is considered in the cylindrical coordinate system ($r , \theta, x$) centred at the axis of the jet with the $x$-axis pointing in the streamwise direction, as depicted in Fig.~\ref{fig:shear_layer_model1}. Assuming homogeneity of the flow in the streamwise, azimuthal directions and in time, pressure fluctuations outside and inside of the jet and the radial displacement of the shear-layer take the form,
\begin{equation}
    \left[\begin{array}{c}
p_{\mathrm{ext}}(x,r,\theta,t) \\
p_{\mathrm{int}}(x,r,\theta,t) \\
\zeta(x,\theta,t)
\end{array}\right]=\left[\begin{array}{c}
\hat{p}_{\mathrm{ext}}(r) \\
\hat{p}_{\mathrm{int}}(r) \\
\hat{\zeta}
\end{array}\right] e^{i(kx+m \theta -\omega t)}
\label{eq:p_prime_theta}
\end{equation}
where the subscripts int and ext refer to the small-amplitude disturbances superimposed on the mean flow inside and outside the jet,  $k$ is the streamwise wavenumber, $m$ is the azimuthal wavenumber, $\omega$ is the angular frequency, $p$ is the pressure, $\zeta$ represents the vertical displacement of the shear layer and $\widehat{~.~}$ denotes the Fourier transform. Substituting the pressure fluctuations inside and outside the jet from Eq.(\ref{eq:p_prime_theta}) in the linearised continuity, momentum and energy equations of a compressible inviscid fluid leads to the following set of equations: 
\begin{equation}
~~~~~~~~~~~~~~~\frac{\partial^{2} p_{\mathrm{ext}}}{\partial t^{2}} -a_{0}^{2} (\frac{1}{r^{2}}\frac{\partial^{2} p_{\mathrm{ext}}}{\partial^{2} \theta}+\frac{\partial^{2} p_{\mathrm{ext}}}{\partial^{2} x})=0 \text{~~~outside of the jet,}
\label{eq:out_jet}
\end{equation}
\begin{equation}
\frac{\partial  p_{\mathrm{int}}}{\partial t^{2}}+U_{j}^{2} \frac{\partial p_{\mathrm{int}} }{\partial x^{2}}-a_{j}^{2}  (\frac{1}{r^{2}}\frac{\partial^{2} p_{\mathrm{int}}}{\partial^{2} \theta}+\frac{\partial^{2} p_{\mathrm{int}}}{\partial^{2} x})=0 \text{~~~inside of the jet,}
\label{eq:inside_jet}
\end{equation}
and the compatibility conditions at the interface between the jet region and region at rest ($r/D=\pm 0.5$):
\begin{equation}
p_{\mathrm{int}} =p_{\mathrm{ext}}; ~~~~~\frac{-1}{\rho_{0}} \frac{\partial p_{\mathrm{ext}}}{\partial r} =\frac{\partial^{2} \zeta}{\partial t^{2}}; ~~~~~-\frac{1}{\rho_{j}} \frac{\partial p_{\mathrm{int}}}{\partial r} =\frac{\partial  \zeta}{\partial t^{2}}+U_{j}^{2} \frac{\partial \zeta }{\partial x^{2}}
\end{equation}
where $\rho$, $U_{j}$  are the density and jet velocity,  $a_{0}$ and $a_{j}$ are the speed of sound in the field at rest and in the jet respectively. The mathematical solution  of Eq.(\ref{eq:out_jet}) that satisfies the outgoing wave or boundedness condition as $r \to \infty$ involves two square-root  functions in the complex $k$-plane.  Square-root  functions  are  multi-valued  functions. As such, they are not allowed in a physical solution unless they are first made single-valued mathematically. In order to obtain such single-valued solutions, branch points and branch cuts need to be inserted in the complex $k$-plane to ensure that boundary conditions are satisfied (boundedness of the eigenfunctions) for any point $k$ in the complex $k$-plane and also to make sure that the branch cuts do not interfere with the inverse $k$-contour. For the pressure disturbance distribution outside of the jet, the solution is explicitly provided with the corresponding choice of branch. Inside of the jet, the equation is solved numerically, but a careful choice of branch must be done, as detailed in the works from Tam and Burton~\cite{Tam1984}. To ensure single-valued solutions, the branch cuts of $\eta_{\mathrm{ext}}$ and $\eta_{\mathrm{int}}$ are
taken to be (Tam and Hu~(1989)~\cite{Tam1989}):
\begin{equation}
-\frac{1}{2} \pi < arg(\eta_{\mathrm{ext}}) \leqslant \frac{1}{2} \pi,~~~~ 0 \leqslant arg(\eta_{\mathrm{int}}) < \pi
\end{equation}
where $\eta_{\mathrm{ext}}=(k^{2}-\omega^{2}/a_{0}^{2})^{1/2}$ and $\eta_{\mathrm{int}}=(( \omega-U_{j} k)^{2}/U_{j}^{2}-k^{2})^{1/2}$. Tam and Hu (1989)~\cite{Tam1989} obtained the following solution for the pressure disturbances in the jet and dispersion relation (DPR):
\begin{equation}
\hat{p}_{\mathrm{int}}=H_{m}^{(1)}(i \eta_{\mathrm{ext}} \frac{D}{2}) \frac{J_{m}(\eta_{\mathrm{int}} r)}{J_{m}(\eta_{\mathrm{int}} \displaystyle{\frac{D}{2})}}
\label{eq:dist_round}
\end{equation}
\begin{equation}
DPR(\omega, k) \equiv \frac{\mathrm{i} \eta_{\mathrm{ext}}}{\rho_{\mathrm{o}} \omega^{2}} J_{m}\left(\eta_{\mathrm{int}} \frac{D}{2} \right) H_{m}^{(1) \prime}\left(\mathrm{i} \eta_{\mathrm{ext}} \frac{D}{2} \right)-\frac{\eta_{\mathrm{int}}}{\rho_{j}\left(\omega-U_{j} k\right)^{2}} H_{m}^{(1)}\left(\mathrm{i} \eta_{\mathrm{ext}} \frac{D}{2} \right) J_{m}^{\prime}\left(\eta_{\mathrm{int}} \frac{D}{2} \right)=0
\label{eq:disp_round}
\end{equation}
where $J_{m}$( ) is the Bessel function of order $m$ and $H_{m}^{(1)}$( ) is the $m$-th order Hankel function of the first kind, ()' denotes the derivative. It can be noted that $m=0$ refers to the first axisymmetric mode and $m=1$ to the first azimuthal mode (helical). Eigenmodes of the vortex-sheet model are defined as frequency-wavenumber pairs that satisfy equation \ref{eq:disp_round} for a given $m$. In order to find dispersion relations for the guided jet modes, we set real values for $\omega$ and find the associated values of $k$.

\subsection{Locally-parallel stability model}
\begin{figure}
\centering
\includegraphics[trim=0cm 0cm 0cm 0cm, clip=true,width=0.5\linewidth]{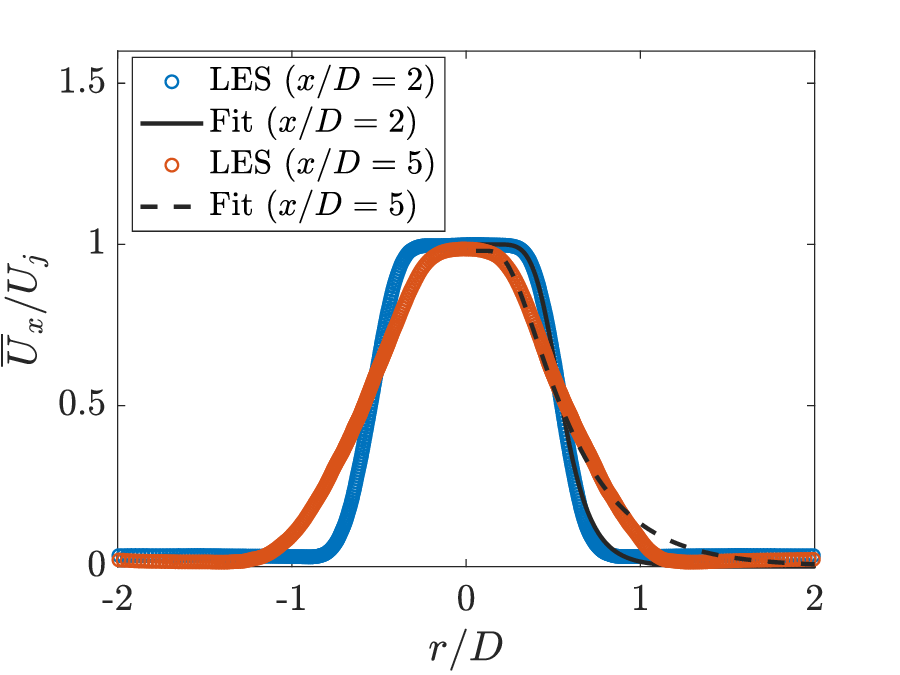}
\caption{Example of mean flow fits. Circles correspond to the mean numerical profiles, computed at $x/D=2$ and $x/D=5$, and the solid and dashed lines correspond to fitted profiles.}
\label{meanfit}
\end{figure}

The stability model considers the compressible Euler equations for flow fluctuations (in a Reynolds-decomposition sense), linearised about the mean flow. Under the assumption of parallel flow, all streamwise derivatives of mean quantities are neglected, $\partial \bar{()}/\partial x = 0$. Using the axisymmetry of the jet, we can also set $\bar{\mathbf{u}} = \left[ \bar{U}_{x}(r), \bar{U}_{r}=0, \bar{U}_{\theta}=0 \right]$, $\bar{\rho}=\bar{\rho}(r)$ and $\bar{T}=\bar{T}(r)$. The density and temperature profiles are obtained based on the mean velocity profile using the Crocco-Busemann relation. Furthermore, the same normal mode \textit{Ansatz} used for the pressure in the previous vortex sheet model is assumed for all flow variables, leading to the following system of equations:

\begin{equation}
    \begin{aligned}
        -i\omega\hat{\rho} + \overline{U}_xik\hat{\rho} +\frac{\partial \overline{\rho}}{\partial r}\hat{u}_r + \overline{\rho} \left( ik\hat{u}_x + \frac{\partial \hat{u}_r}{\partial r} + \frac{\hat{u}_r}{r} + \frac{im\hat{u}_{\theta}}{r} \right) = 0,\\
        \overline{\rho} \left(-i\omega\hat{u}_x + ik\overline{U}_x \hat{u}_x + \frac{\partial \overline{U}_x}{\partial r} \hat{u}_r\right) = -\frac{1}{M_j\gamma}\left( i\overline{\rho}k\hat{T} + i\overline{T} k \hat{\rho} \right),\\
        \overline{\rho}\left( -i\omega\hat{u}_r + \overline{U}_xik\hat{u}_r \right) = -\frac{1}{M_j\gamma}\left( \overline{\rho}\frac{\partial \hat{T}}{\partial r} + \hat{T}\frac{\partial \overline{\rho}}{\partial r} + \overline{T}\frac{\partial \hat{\rho}}{\partial r} \hat{\rho}\frac{\partial \overline{T}}{\partial r} \right),\\
        \overline{\rho}\left( -i\omega \hat{u}_{\theta} + \overline{U}_xik\hat{u}_{\theta} \right) = -\frac{1}{M_j\gamma r} \left( \overline{\rho} im \hat{T} + \overline{T}im\hat{\rho}\right),\\
        \overline{\rho}\left[ -i\omega \hat{T} + \overline{U}_xik\hat{T} + \hat{u}_r\frac{\partial \overline{T}}{\partial r} + (\gamma-1)\overline{T} \left( ik\hat{u}_x + \frac{\partial \hat{u}_r}{\partial r} + \frac{\hat{u}_r}{r} + \frac{im \hat{u}_{\theta}}{r} \right) \right] = 0,
    \end{aligned}
\end{equation}
where $\gamma$ is the specific heat ratio. We solve a spatial stability problem, wherein a real $\omega$ is set and the system is cast into the form of an eigenvalue problem,

\begin{equation}
(-\omega\mathbf{I} +  \mathcal{A}_{0})\hat{\mathbf{q}}=k\mathcal{A}_{1}\hat{\mathbf{q}},
\label{eigen_spatial}
\end{equation}
where $\mathcal{A}_{0}$ and $\mathcal{A}_{1}$ are matrices that group $k^0$ (terms not dependent on $k$) and $k^1$ terms (depending linearly on $k$), respectively. The radial direction is discretised using Chebyshev collocation points. We found that 500 points were sufficient to attain converged eigenvalue spectra. The domain is extended to the far-field by mapping the original domain,  $r \in [-1, 1]$ to $r \in [0,\infty[$ using a function proposed by \citet{trefethen2000spectral} that concentrates most points in the sheared region. The boundary conditions are dependent on the azimuthal wavenumber, as described in \citet{Maia_PRF2023}. Once the eigenvalue problem is solved, pressure eigenfunctions can be reconstructed through the ideal gas state equation using the mean flow, density and temperature perturbations,

\begin{equation}
\hat{p} = \frac{\overline{\rho}\hat{T}+\overline{T}\hat{\rho}}{\gamma M_j^2}.   
\end{equation}
The mean flow profiles that served as input to the model were based on the numerical data, and fitted with the hyperbolic tangent profile from \citet{MICHALKE1984159},

\begin{equation}
    \overline{U}_x = 0.5 \left[1 +\tanh{\left( \frac{R}{4\theta_R} \left( \frac{R}{r} - \frac{r}{R} \right) \right)} \right]
    \label{michalke_mean}
\end{equation}
where $\theta_R$ is the shear-layer thickness. The latter is determined from a least-squares fit of the computed mean flows. Figure \ref{meanfit} shows examples of fits based on the mean profiles computed at $x/D=2$ and $x/D=5$. These are the positions where stability analysis are carried out for cases $L=3D$ and $L=6D$, as will be explained shortly. The analytical function is seen to provide a very good fit for the data at $x/D=2$. At $x/D=5$, a slight mismatch occurs in the outer region. This, however, has negligible impact on the stability results, since the region of high shear is well represented by the analytical profile.

%As explained above, we aim at assessing whether the tones observed in the literature are consistent with the frequency bands where upstream-travelling guided jet modes are neutrally stable (propagative). Such modes belong to a family of waves organised by azimuthal wavenumber, $m$ and radial order, $n$. The modes are propagative, and thus eligible to close the feedback loop, only in certain bands defined by saddle and branch points of the stability spectrum. 

\begin{figure}
\centering
\includegraphics[trim=3cm 11cm 3cm 11cm, clip=true,width=\linewidth]{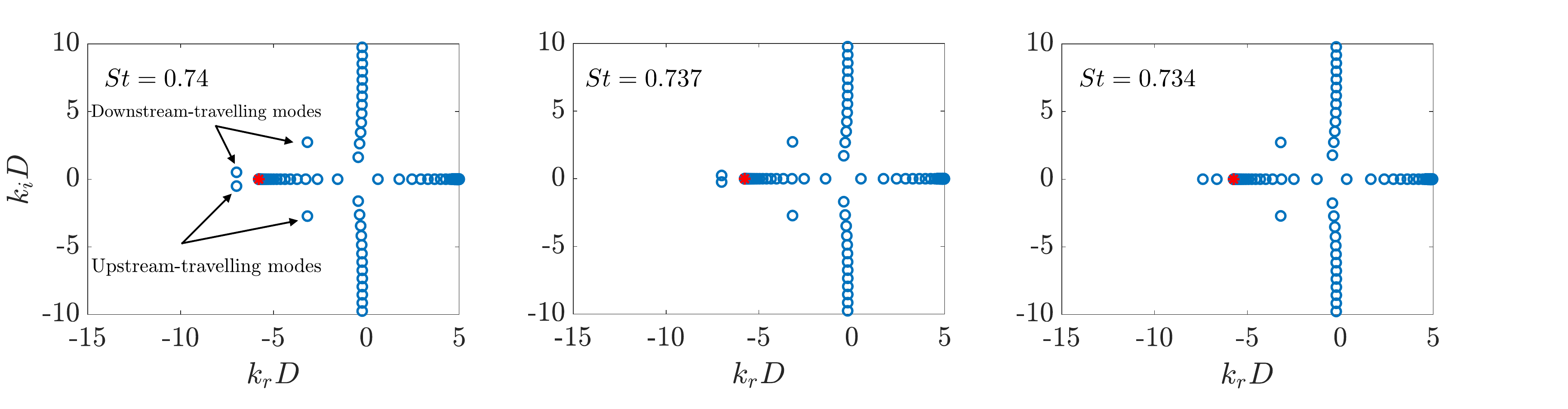}
\caption{Local stability spectra for the axisymmetric $m=0$ mode at different Strouhal numbers in the vicinity of a saddle point. Computations correspond to a velocity profile with $R/\theta_R = 5$. $k_r$ and $k_i$ are the real and imaginary parts of the complex streamwise wavenumber. The red asterisk marks the branch point.}
\label{spectra_punk}
\end{figure}

Unlike the previous model, which is solved directly for neutral modes, solution of the eigenvalue problem given by equation \ref{eigen_spatial} yields the entire stability spectrum. The characteristics of such spectra, and those of the main mechanisms of hydrodynamic and acoustic nature associated with it, have been extensively discussed previously \cite{RodriguezEuropean}. Figure \ref{spectra_punk} shows examples of stability eigenspectra computed for $m=0$ and velocity profile with $R/\theta_R=5$, at Strouhal numbers in the vicinity of a saddle point. The spectra possess four branches: two branches lying on the real axis, consisting of acoustic modes propagating downstream and upstream; and two branches of evanescent
modes, one with positive group velocity and another with negative group velocity. Two sets of discrete modes can be seen in (a)-(c), belonging to two different radial orders (which was determined by inspection of the number of anti-nodes in their respective eigenfunctions). One of these sets has negative group velocity (upstream-travelling) and the other has positive group velocity (downstream-travelling). The sign of the group velocity, $\partial \omega/\partial k_r$, (with $k_r$ the real part of the streamwise wavenumber) can be determined by a complex-frequency analysis; this is not shown here for conciseness. Waves belonging to the former group are known as \textit{guided jet modes}, and are those involved in the aeroacoustic loop. The downstream-travelling discrete modes are known as duct-like modes, given their similarity to acoustic modes of a cylindrical soft duct, as discussed in \citet{towne_cavalieri_jordan_colonius_schmidt_jaunet_brès_2017}. We have verified that their eigenfunctions indeed correspond to those issuing from eigenanalysis of a soft-walled duct. This is not shown here for brevity. As the Strouhal number decreases from (a)-(c), the modes approach the real $k$ axis. They reach the real axis at the saddle point, at $St \approx 0.736$. With decreasing Strouhal number below this point, the downstream-travelling mode moves toward $- \infty$, whereas the upstream-travelling mode gradually approaches the branch of acoustic modes, until it finally reaches the branch point, which is marked in red in the figure. 

The dispersion relations of such modes for different radial orders are then constructed by varying $St$ and tracking the real part of their wavenumbers as they become propagative ($k_i=0$).

\section{Results \label{sec:results}}

\begin{figure}
\centering
\includegraphics[trim=5cm 3cm 5cm 3cm, clip=true,width=\linewidth]{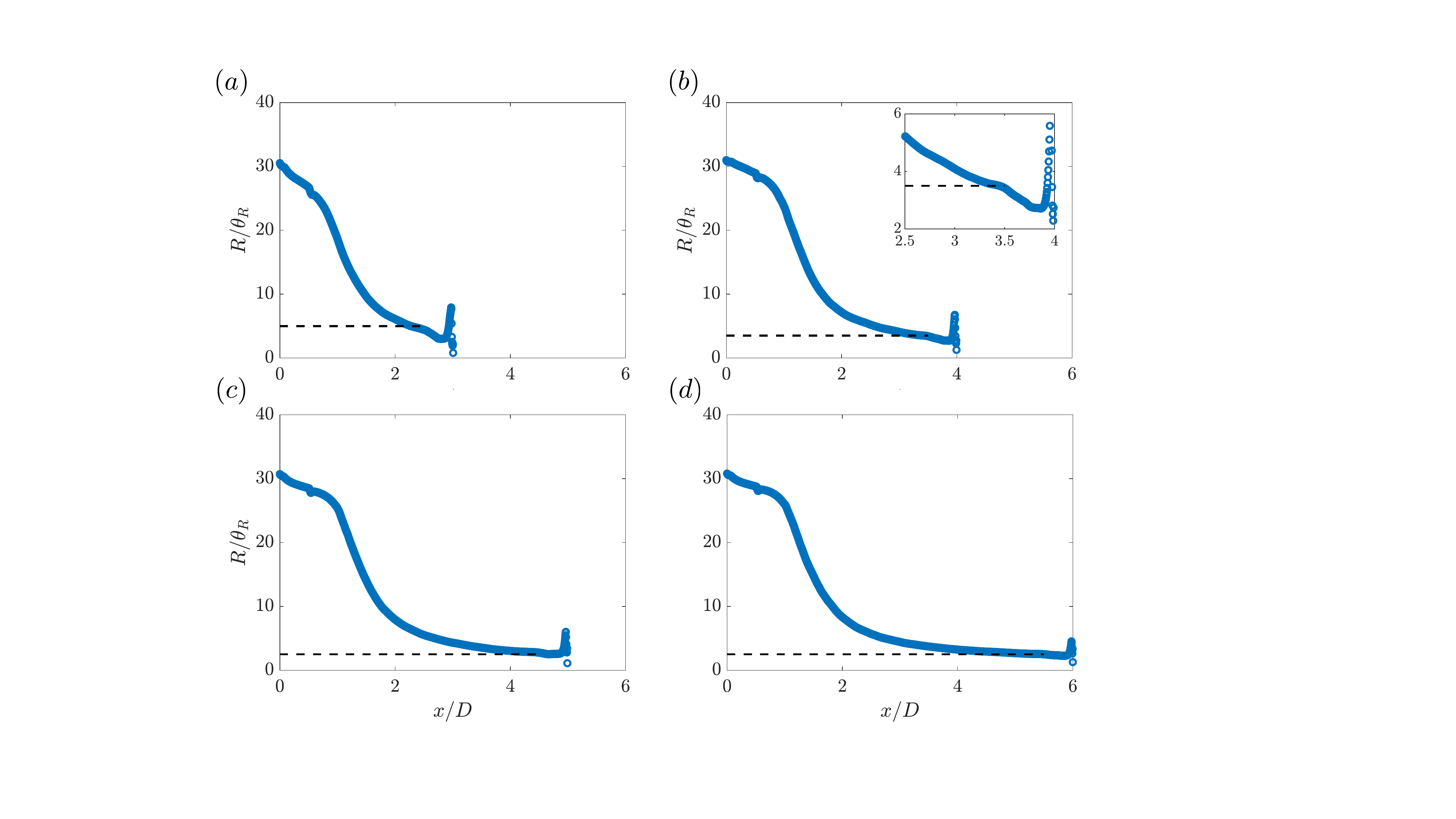}
\caption{Streamwise evolution of the shear-layer thickness in the four simulated cases: (a) Jetideal3D; (b) Jetideal4D; (c) Jetideal5D; (d) Jetideal6D. The inset in (b) displays a zoom on the vicinity of the plate. Black dashed lines represent the values of $R/\theta_R$ considered for each case.}
\label{r_theta}
\end{figure}

In the following, we compare dispersion relations of the different models and we assess to what extent they are consistent with the tones observed in the LES database. For the finite-thickness models, the velocity profile close to the reflection point is necessary. While in screeching jets the reflection point is often located between the fourth and the sixth shock cell  \cite{gojon2017numerical,mancinelli2019screech, mancinelli_jaunet_jordan_towne_2021}, for ideally-expanded impinging jets the natural choice is to consider the reflection to occur at the plate. Figure \ref{r_theta} displays the streamwise evolution of the shear-layer thickness for the four simulated cases. In the close vicinity of the plate, the mean profile is significantly distorted, as can be seen in the inset of figure \ref{r_theta}(b). There is a sudden change in the slope of the $R/\theta_R$ parameter around $1D$ upstream of the plate, followed by a sharp increase in its immediate vicinity. This is accompanied by a strong decay of the jet centerline velocity (not shown). In order to avoid strong mean flow distortion effects on the stability characteristics, we consider shear-layer thicknesses computed $1.0-1.2D$ upstream of the plate. This leads to the values $R/\theta_R = 5, 3.5, 2.5$, and $2.5$ for the cases Jetideal3D, Jetideal4D, Jetideal5D and Jetideal6D respectively.

\begin{figure}
\centering
\includegraphics[trim=10cm 10cm 10cm 10cm, clip=true,width=\linewidth]{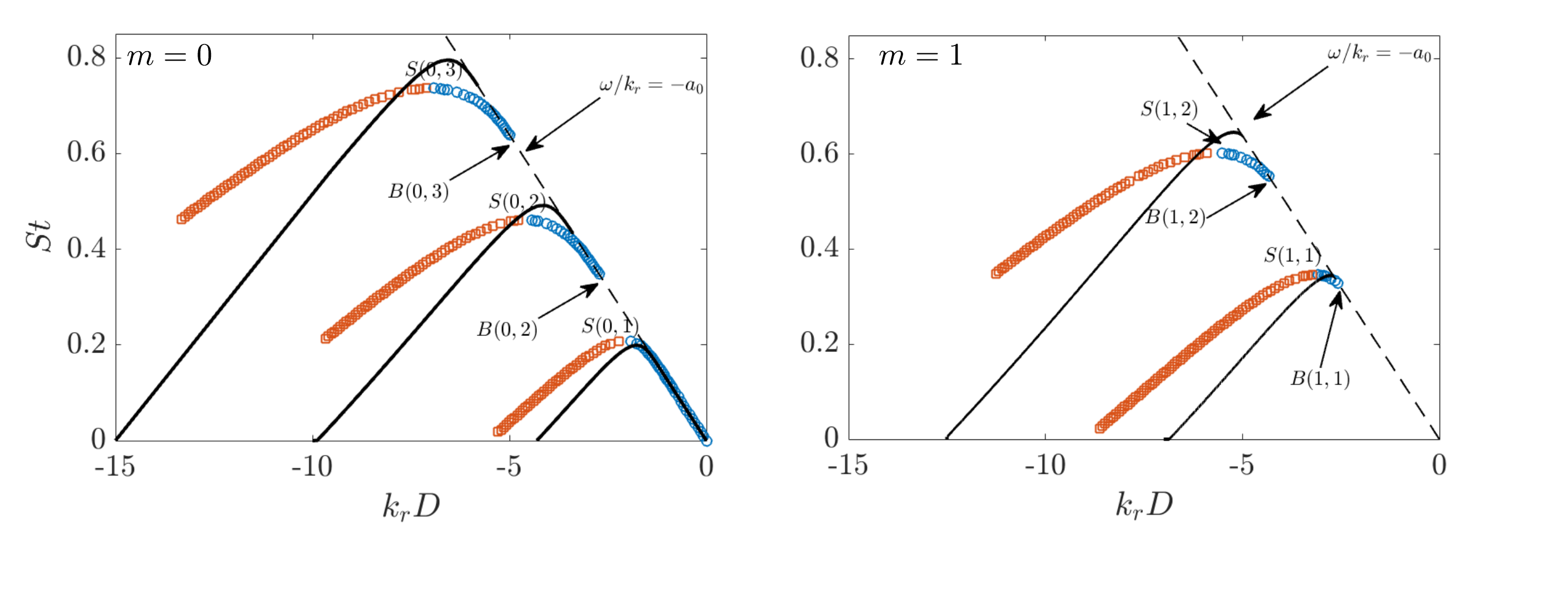}
\caption{Dispersion relations of the upstream-travelling (blue circles) and downstream-travelling (red squares) guided jet modes for azimuthal modes $m=0$ and $m=1$ and $R/\theta_R=5$. Solid black lines are vortex-sheet solutions based on Eq.~(\ref{eq:disp_round}). Saddle and branch points of different radial orders (denote n) for the finite-thickness model are indicated by the $S(m,n)$ and $B(m,n)$, respectively. The diagonal dashed line marks the sonic line.}
\label{disp_rel}
\end{figure}

Figure \ref{disp_rel} shows examples of dispersion relations of the guided jet waves, computed for azimuthal wavenumbers $m=0,1$ and $R/\theta_R = 5$. Downstream-travelling modes are represented by the red squares and upstream-travelling modes by blue circles. Notice that this distinction is based on the sign of their group velocity, $\partial \omega / \partial k$, which can be inferred from the slope of the curves in the $k_r-St$ plane. In this Strouhal number range, neutral modes of three different radial orders exist for the axisymmetric mode, while two different radial orders are visible for $m=1$. The n number (n=1, 2,3,...) is the radial mode number characterizing the number of anti-nodes (maximum oscillation points) of the pressure distribution of the wave
in the radial direction (visible for example in the eigenvectors). The vortex-sheet solution is also displayed for comparison. Inclusion of the shear-layer thickness on stability model clearly modifies the dispersion relations of the upstream-travelling modes. Two main effects are noticeable: a decrease in the Strouhal numbers of the saddle and branch points; and a widening of the Strouhal number range where these modes are propagative.

\begin{figure}[!h]
\centering
\includegraphics[trim=10cm 0cm 5cm 0cm, clip=true,width=\linewidth]{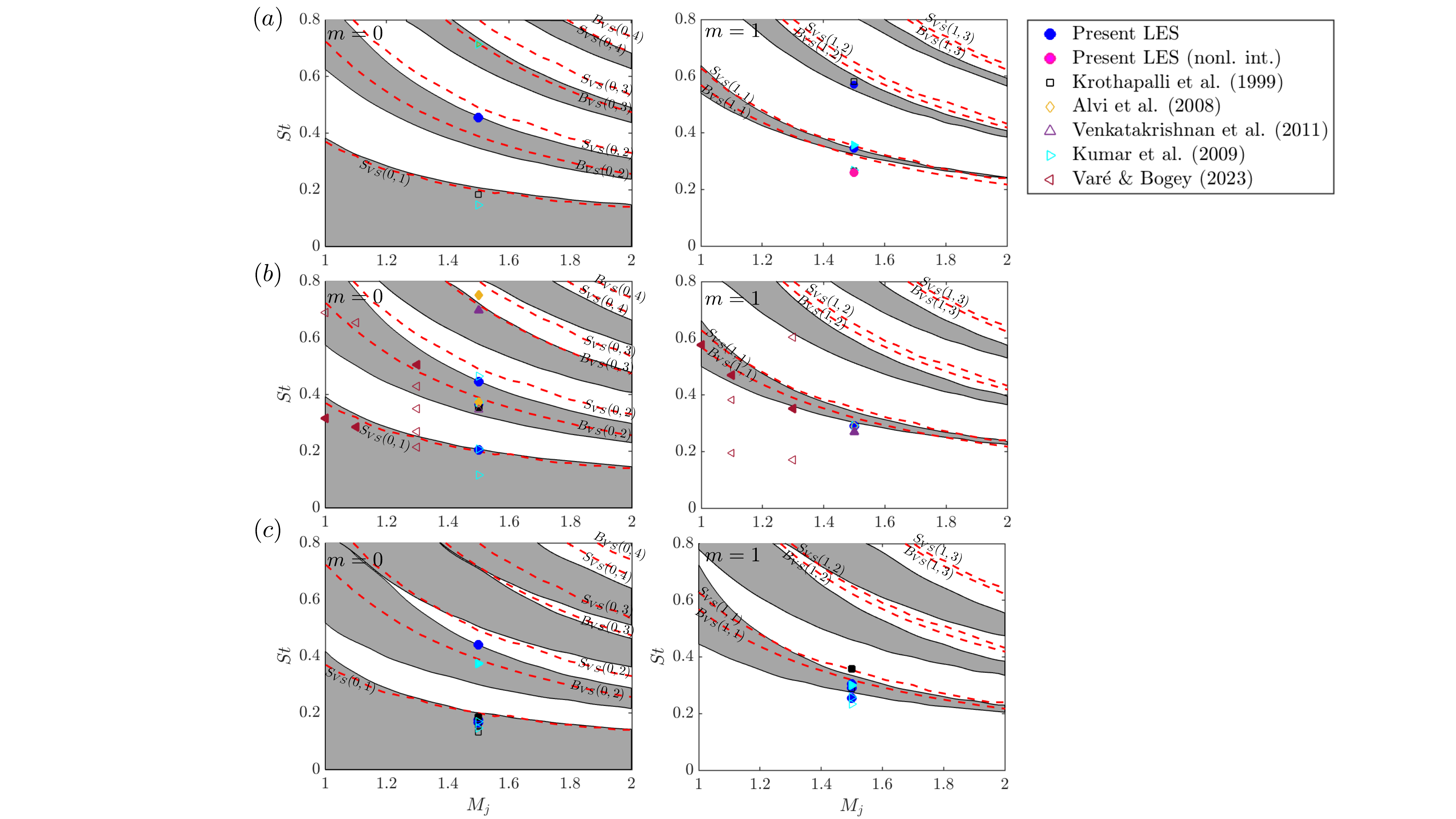}
\caption{Allowable frequency ranges for resonance, delimited by the saddle and branch points of the dispersion relation. Gray-shaded regions are frequency ranges predicted by the finite-thickness stability model, and the dashed red lines correspond to the limits of the vortex-sheet model. The branch and saddle points of the vortex sheet model for different azimuthal wavenumbers and radial orders, which delimit the frequency ranges, are indicated by $S_{VS}$ and $B_{VS}$, respectively. In (a), (b) and (c), the limits of the finite-thickness stability model are computed for $R/\theta_{R}=5$, $R/\theta_{R}=3.5$ and $R/\theta_{R}=2.5$, respectively. The blue circles correspond to tones observed in the LES and the tone denoted by the pink circle in (a) is the result of a nonlinear interaction between an axisymmetric and a helical mode, as demonstrated in Appendix A. The other symbols correspond to experimental and numerical points found in the literature for the same nozzle-to-plate distances used in the LES. The dominant tone (or the two most dominant tones, when available) in each database, are indicated with filled symbols.}
\label{tones}
\end{figure}

These dispersion relations were used to compute allowable frequency ranges for resonance, delimited by the saddle and branch points of different radial orders. They are presented in figure \ref{tones} for jet Mach numbers in the range $M_j=1.0-2.0$ and azimuthal modes $m=0,1$. The tones observed in the LES data at $M_j=1.5$ are represented by blue filled circles.  Cases $L=3D$ and $L=4D$ are modelled with $R/\theta_R=5$ and $R/\theta_R=3.5$, respectively, following the results shown in figure \ref{r_theta}, and are shown in (a) and (b). Tones observed for $L=5D$ and $6D$, on the other hand, are displayed together in (c), since the shear-layer thickness was found to be almost the same in both cases ($R/\theta_R=2.5$). In order to make the assessment of the model as extensive as possible, tonal frequencies observed in other experimental \cite{krothapalli_rajkuperan_alvi_lourenco_1999, alvi2008experimental, venkatakrishnan2011density, kumar2009control} and numerical \cite{vare_bogey_aiaa2023} studies of ideally-expanded round impinging jets are also shown for comparison. These databases cover the same range of nozzle-to-plate distances used in the present study, and the experimental/numerical data points are compared to predictions made with $R/\theta_R=5, 3.5, 2.5$, following the same criterion described above for the current LES.

Some considerations must be made regarding the comparison with data from previous studies. First of all, the underlying hypothesis in this comparison is that, if the nozzle-to-plate distance is the same, the shear-layer thickness at the reflection point is also the same. This might not be strictly true, as the jets issue from different nozzle geometries, with possibly different jet Reynolds numbers. However, the vast majority of these studies were conducted at $M_j=1.5$ and display tones whose frequencies are fairly close to each other, and close to those observed in our database. This suggests that the shear-layers have comparable thicknesses. Secondly, the frequency ranges of existence of upstream-travelling modes are sensitive to the jet temperature \cite{mancinelli_jaunet_jordan_towne_2021}. Therefore, this comparison is only possible because in all these studies the jet stagnation temperature is quite similar. Finally, in the experimental results of studies \cite{krothapalli_rajkuperan_alvi_lourenco_1999, alvi2008experimental, venkatakrishnan2011density, kumar2009control}, the pressure field is not Fourier-decomposed in azimuth as in the present study (and also in the numerical study of \citet{vare_bogey_aiaa2023}), and the azimuthal nature of the tones (axisymmetric or helical) is not indicated. Therefore, in order to perform the comparison with the stability models, we have made an \textit{ad-hoc} separation of the experimental data points into $m=0$ or $m=1$ tones according to their proximity to the LES data: if the frequency of the experimental tone is close to a $m=0$ LES tone, it is assigned as axisymmetric; if it is close to a $m=1$ LES tone, it is assigned as helical. The justification for this assumption comes from the similarity in the jet exit conditions and nozzle-to-plate distances, as discussed above. A few experimental points were found to be quite far from the LES data. In that case, they have been assigned a wavenumber in a more arbitrary way, based on their proximity to the vortex-sheet solution. Additionally, we have also tried to split all the experimental datapoint based on their agreement with the vortex-sheet solutions. Apart from minor differences, the trends are quite similar to those shown in figure \ref{tones}, and are not shown here for conciseness.

Figure \ref{tones} reveals that the discrepancy between the vortex-sheet and finite thickness models clearly becomes more pronounced as the shear-layer thickness is increased. The allowable frequency ranges are gradually widened with decreasing $R/\theta_R$, and the saddle and branch points are progressively shifted to lower Strouhal numbers. Many of the observed tones are inconsistent with the allowable frequency ranges of the vortex-sheet model. But with the inclusion of a finite thickness consistent with the reflection location, all the LES points are either well within the limits predicted by the model or very close to them. The tone represented by the pink dot in (a), which is well off the allowable ranges for any of the models, was found to result from a nonlinear interaction between an axisymmetric and a helical tone, as shown in Appendix A via a bispectral mode decomposition \cite{schmidt2020bispectral}. It is therefore not generated by a fundamental instability mechanism. The vast majority of experimental and numerical points from the literature are also more consistent with the finite thickness model. There are some exceptions, notably the $m=1$ non-dominant peaks reported by \citet{vare_bogey_aiaa2023}, which are well off the model predictions. One possible explanation for this mismatch is that these tones also be generated by nonlinear interactions. But apart from the few discrepancies, overall, the observed trends indicate that accounting for a finite thickness shear-layer is necessary for accurate tone predictions in impinging jets, analogous to what was recently observed for the screech phenomenon \cite{mancinelli_jaunet_jordan_towne_2021}.

%\begin{figure}
%\centering
%\includegraphics[trim=8cm 1cm 8cm 1cm, %clip=true,width=0.8\linewidth]{Figures/wavenumber_frequency_diagram.pdf}
%\caption{Frequency-wavenumber diagrams of pressure fluctuations extracted at jet centerline, $r/D=0$. From \cite{ferreira_aiaa2023}}
%\label{wave_freq_spec}
%\end{figure}

\begin{figure}
\centering
\includegraphics[trim=3cm 14cm 7cm 8cm, clip=true,width=\linewidth]{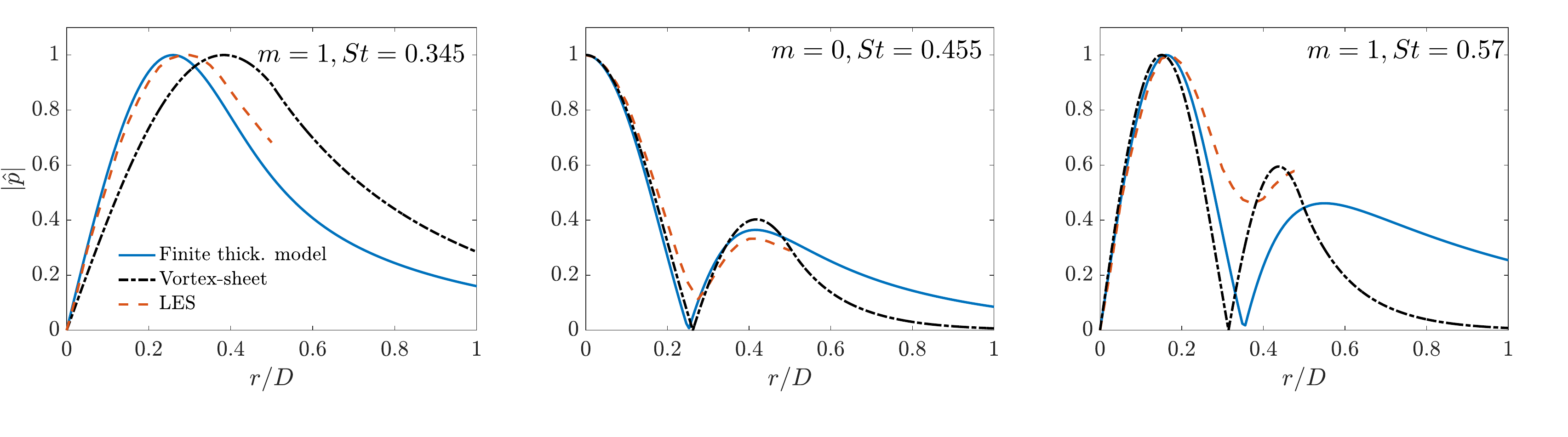}
\caption{Comparison between stability eigenfunctions of the upstream-travelling jet mode (solid blue lines), vortex sheet model based on Eq.~(\ref{eq:dist_round}) (dashed-dot black lines) and LES pressure data (dashed red lines), computed at frequencies corresponding to the three dominant peaks identified in the case JetIdeal3D. The curves are normalised by their maximum values. }
\label{comp_eigenfunctions}
\end{figure}

Apart from the tone frequency, the radial structure of the modelled waves was also compared to that of upstream-travelling perturbations issuing from the LES. The latter are educed following the procedure laid out by \citet{ferreira_aiaa2023}, wherein the pressure field is Fourier transformed in time and in the streamwise direction at different radial positions between $r/D=0-0.5$, producing frequency-wavenumber diagrams. The radial structure of the upstream-travelling waves is then extracted by averaging, at each radial position, the Fourier coefficients of a few streamwise wavenumbers around the sonic line ($\omega/k=a_0$), at the tone frequency. This averaging is done because the wavenumber resolution of the Fourier transform is quite poor, due to the short nozzle-to-plate distance that sets the wavenumber discretisation. Therefore, averaging over a few wavenumbes around the sonic line helps obtaining smoother profiles. Notice that the contribution from the wavenumber predicted from linear stability theory is contained in that average. The radial profiles educed for the JetIdeal3D database are compared in figure \ref{comp_eigenfunctions} to eigenfunctions of the vortex-sheet and stability models, computed at the Strouhal numbers of three dominant peaks. For the tones at $St=0.455$ and $St=0.57$, the eigenfunctions of both models are in reasonable agreement with the LES. For the first helical tone at $St=0.345$, the eigenfunction from the finite-thickness model provides a much better agreement with the data, which further reinforces the necessity of taking the shear layer structure into account for an accurate modelling of the guided jet modes. Slight mismatches occur, which is expected, due to the averaging procedure mentioned above. The strongest mismatch is around the dip (at $r/D=0.38$) in the structure of the $St=0.57$ tone. Such dips result from phase jumps in vortical structures and are difficult to observe correctly in the data due to the stochastic background turbulence that distorts those structures. This mismatch can be corrected, to a great extent, by extracting the organised part of the pressure field using proper orthogonal decomposition, \cite{Cavalieri2013}. This is, however, outside the scope of this work. But overall, the agreement between models and data is reasonable, which provides further evidence that upstream-travelling guided jet waves are indeed involved in the resonance mechanism.

The present work follows previous modelling studies that consider locally-parallel frameworks to model guided jet modes \cite{gojon2018oscillation, edgington-mitchell_jaunet_jordan_towne_soria_honnery_2018, mancinelli2019screech,mancinelli_jaunet_jordan_towne_2021, jaunet_aiaa2019, vare_bogey_aiaa2023}. This framework provides good approximations for these waves, even when shocks are present, which allows it to be used to predict tone frequencies in screeching jets. Furthermore, it has been shown in the work of \citet{mancinelli_jaunet_jordan_towne_2021} that, in order to accurately predict frequency tones at different Mach numbers, the finite-thickness of the shear layer \textit{at the downstream reflection point} should be taken into account, which clearly points to an advantage of finite-thickness stability models over the classic vortex-sheet model. At the point where the guided jet wave is excited (the downstream reflection point), it should exist as a neutral discrete mode (or, at the very least, as an evanescent wave), otherwise the resonant loop is not possible. But in most cases, when an infinitely thin shear layer is considered at that location, the guided jet mode does not exist as a discrete wave at the resonance frequency. This can be clearly seen in the results of the present study. Notice that several tones shown in figure \ref{tones} are below the branch point of the vortex-sheet, which means that, at the tone frequency, they are absorbed in the continuous branch of acoustic free-stream modes. Accounting for the shear layer thickness allows us to reconcile the guided jet waves with the observed tones, which, in in our view, highlights the importance of correctly modelling the jet dynamics at the downstream reflection point.

However, the following consideration should be made: as the guided jet mode travels upstream, it encounters increasingly thinner shear layers. When it reaches the upstream reflection point (the nozzle exit) the shear layer is quite thin, and if we choose this position to compute the characteristics of the guided modes, the results of the finite thickness model would then be quite similar to those of the vortex sheet model, leading to mismatches with the data, as observed in previous studies. This apparent conundrum seems to indicate that the resonance frequencies are governed rather by the downstream reflection point. This remains an open question for the moment, but we can make the following conjecture: in order for the guided jet mode not to lose its support when travelling upstream, and eventually close the loop at the nozzle, it must \enquote{see} an average shear-layer between the plate and the nozzle whose thickness is compatible with its existence. In Appendix B we explore the validity of this hypothesis by considering a streamwise-averaged mean flow of the $L=3D$ case. The results show that the frequency bands of waves associated with the spatially-averaged flow are also close to the tones observed. This suggests that this \enquote{average shear layer} could be considered as a surrogate for for the shear layers at the upstream and downstream reflection locations, avoiding the choice of one or the other to perform tone predictions. Of course this streamwise-averaged flow scenario is just a rough approximation of the dynamics of these waves between the nozzle and the plate. A more accurate representation of their behaviour as they travel upstream can be obtained with a global stability calculation, as done by \citet{hildebrand2017global}. In the present work we choose to carry out the main analyses at the downstream reflection position, for consistency with recent works that consider finite thickness models for screeching jets  \cite{mancinelli_jaunet_jordan_towne_2021}.

\section{Conclusions \label{sec:conclusions}} 

The origin of resonance tones in ideally-expanded round impinging jets has been investigated. We hypothesise that resonance is underpinned by a downstream-travelling Kelvin-Helmholtz instability and upstream-travelling guided jet modes, as done in previous studies \cite{gojon_bogey_marsden_2016, Bogey2017, jaunet_aiaa2019, varé_bogey_2022, vare_bogey_aiaa2023, ferreira_aiaa2023}. In these studies, the allowable frequency bands for resonance, dictated by the propagative character of the upstream-travelling waves, are predicted with vortex-sheet models. Vortex-sheet predictions were able to partially explain a number of observed tones; but many tones lied outside the allowable frequency bands predicted by the model \cite{Bogey2017, ferreira_aiaa2023}, and the origin of the discrepancy remained unexplained.

Motivated by a recent study that revealed that accurate tone-frequency predictions in screeching jets requires accounting for a finite-thickness shear layer \cite{mancinelli_jaunet_jordan_towne_2021}, here we explore finite-thickness models in the prediction of allowable resonance ranges on the impinging jet problem. We revisit LES databases of ideally expanded supersonic jets impinging on flat plates, with different nozzle-to-plate distances \cite{Bogey2017, gojon_bogey_aiaa2018}. Mean-flow information in the vicinity of the plate is used as input to stability models. Finite thickness models predict wider allowable frequency bands for resonance whose upper and lower limits are shifted to lower frequencies. These trends become progressively more pronounced as the shear layer thickens. All of the main tones observed in the LES data are found to fall within, or very close to, the frequency ranges obtained with the finite thickness models, improving on vortex-sheet predictions. That improvement is substantiated by comparison with data from the literature: the finite-thickness model, tuned with the shear-layer thickness educed from the LES, was also found to be consistent with a number of tones observed in past experimental and numerical studies carried out with the same range of nozzle-to-plate distances. Furthermore, the radial structure of the upstream-travelling waves computed with the finite-thickness stability model is found to be in good agreement with upstream-travelling disturbances educed from the present LES data, providing further evidence of the involvement of these waves in the resonance mechanism in ideally expanded impinging jets. 

The advantage of using a locally-parallel framework is that its eigenvalue spectrum is more easily interpretable than that issuing from a global analysis, which makes it easier to identify and track the guided and duct modes in the frequency-wavenumber plane. Also, as mentioned above, the locally-parallel framework has been shown in numerous studies to provide good approximations for these waves in screeching jets. However, the global analysis can be useful to overcome limitations of the local analysis. For instance, it can provide a more complete description of the guided jet mode dynamics as it travels from the downstream to the upstream reflection point. This has the benefit of avoiding the choice of a single position to perform tone predictions, as discussed above. Furthermore, the global framework is also more suitable for shock-containing impinging jets \cite{hildebrand2017global}, as in that case multiple reflection mechanisms (at the plate and the shocks) can coexist, making the local analysis more ambiguous. Future work will address global stability analysis to study tone generation in underexpanded impinging jets.

Finally, the results of the present study can serve as guidance and departure points for more refined tone-prediction models in impinging jets. Future models should consider the true structure of the jet shear-layer in the vicinity of the plate. They should also address the phase-matching criteria between downstream- and upstream-travelling disturbances, which is necessary for exact tone-frequency predictions at different Mach numbers. This requires knowledge of reflection coefficients at the nozzle and plate. This can be achieved, for instance, via Wiener-Hopf techniques \cite{rienstra2007acoustic}, or data-driven approaches that involve projection of flow data into stability eigenbases. \cite{Prinja_2023}.

\section*{Acknowledgements}

This work was performed using resources from GENCI [CCRT-CINES-IDRIS] (Grant 2022-[A0122A07178]) and CALMIP (Project 1425-22A). I. A. M. acknowledges funding from the São Paulo Research Foundation (FAPESP/Brazil) through grant number 2022/06824-4.

\appendix

\section*{Appendix A: Nonlinear interaction of tones}
\label{sec:appendix}
In this section we apply the Bispectral Mode Decomposition (BMD) to the Jetideal3D database, with the purpose of explaining tones that fall outside the allowable frequency range of upstream-travelling modes. BMD is a technique developed to study nonlinear triads interactions and coherent structures associated with them \cite{schmidt2020bispectral}. The method performs a maximization of an integral measure of the bispectrum, which, for a pair of frequencies $f_k$ and $f_l$ takes the form,

\begin{equation}
    b(f_{k},f_{l}) = E\left[ \int_{\Omega} \hat{\mathbf{q}}_{k}^{*} \circ \hat{\mathbf{q}}_{l}^{*} \circ \hat{\mathbf{q}}_{k+l}^{*}\right] = E\left[ \hat{\mathbf{q}}_{k\circ l}^{H} \mathbf{W} \hat{\mathbf{q}}_{k+l}\right] = E\left[ \left < \hat{\mathbf{q}}_{k\circ l}^{H},\hat{\mathbf{q}}_{k+l}\right >\right],
\end{equation}
where $\mathbf{q}$ is a fluctuating field of a given state variable, E[] represents the expected value. The hats $\widehat{.}$ denote Fourier transformation in time, computed with the Welch method, asterisks (.)$^{*}$ denote scalar complex conjugation and (.)$^{H}$ the transpose conjugate operator. $\Omega$ is the spatial domain, and the inner product of the right term is expressed as,

\begin{equation}
\left< \mathbf{q}_{1},\mathbf{q}_{2}\right>=\iint\mathbf{q}_{1}^{*}\mathbf{q}_{2} \mathrm{d}x \mathrm{d}\theta \mathrm{d}r=\mathbf{q}_{1}^{*} \mathbf{W} \mathbf{q}_{2},
\label{eq:inner_product}
\end{equation}
where $\mathbf{W}$ is a matrix containing numerical quadrature weights. $\mathbf{q}$ is a vector containing realisations of state variables. Here $\circ$ denotes a Hadamard product, and the notation

\begin{equation}
\hat{\mathbf{q}}_{k\circ l} \equiv \hat{\mathbf{q}}(\mathbf{x},f_{k}) \circ \hat{\mathbf{q}}(\mathbf{x},f_{l})
\end{equation}
denotes the point-wise product of two realizations of the flow at frequencies $f_{k}$ and $f_{l}$. All Fourier realisations at a given frequency $f_k$ are arranged into a matrix of the form $\hat{\mathbf{Q}}_{{k}}=[\hat{\mathbf{q}}_{{k}}^{(1)} \hat{\mathbf{q}}_{{j}}^{(2)} \cdots \hat{\mathbf{q}}_{{k}}^{(N_{blk})}]$ and a bispectral matrix is defined as,

\begin{equation}
\mathbf{B}(\mathbf{x},f_{k}, f_{l}) \equiv \frac{1}{N_{blk}}{\hat{\mathbf{Q}}_{k \circ l}}\mathbf{W}\hat{\mathbf{Q}}_{k + l},
\end{equation}
where $\hat{\mathbf{Q}}_{ k \circ l}^{H} = \hat{\mathbf{Q}}_{k}^{*} \circ \hat{\mathbf{Q}}_{l}^{*}$. The complex mode bispectrum is obtained by solving the following optimisation problem,

\begin{equation}
\mathbf{a}_{1} = \mathrm{arg \; max}\left| \frac{\mathbf{a}^{H}\mathbf{B}\mathbf{a}}{\mathbf{a}^{H}\mathbf{a}} \right|,
\label{rayleighcoeff}
\end{equation}
where $\mathbf{a}_1$ are optimal expansion coefficients.  This problem is solved by computing the numerical radius, $\lambda_1$ of $\mathbf{B}$.  The reader is referred to \citet{schmidt2020bispectral} for details about the numerical procedure. The maps of $\lambda_1(f_k, f_l)$ illustrate regions of the spectrum where there are significant quadratic phase-coupling between different frequencies, generating a third component, $f_j$, which follows the triad $f_j \equiv f_k + f_l$. The method uses flow data available in the entire domain, as opposed to classic bispectral analysis, which considers point-wise data. BMD also provides sets of modes associated with the formation of a given triad; but here we focus only in the complex bispectrum, $\lambda_1(f_k, f_l)$.

\begin{figure}
\centering
\includegraphics[trim=1cm 8cm 2cm 7cm, clip=true,width=\linewidth]{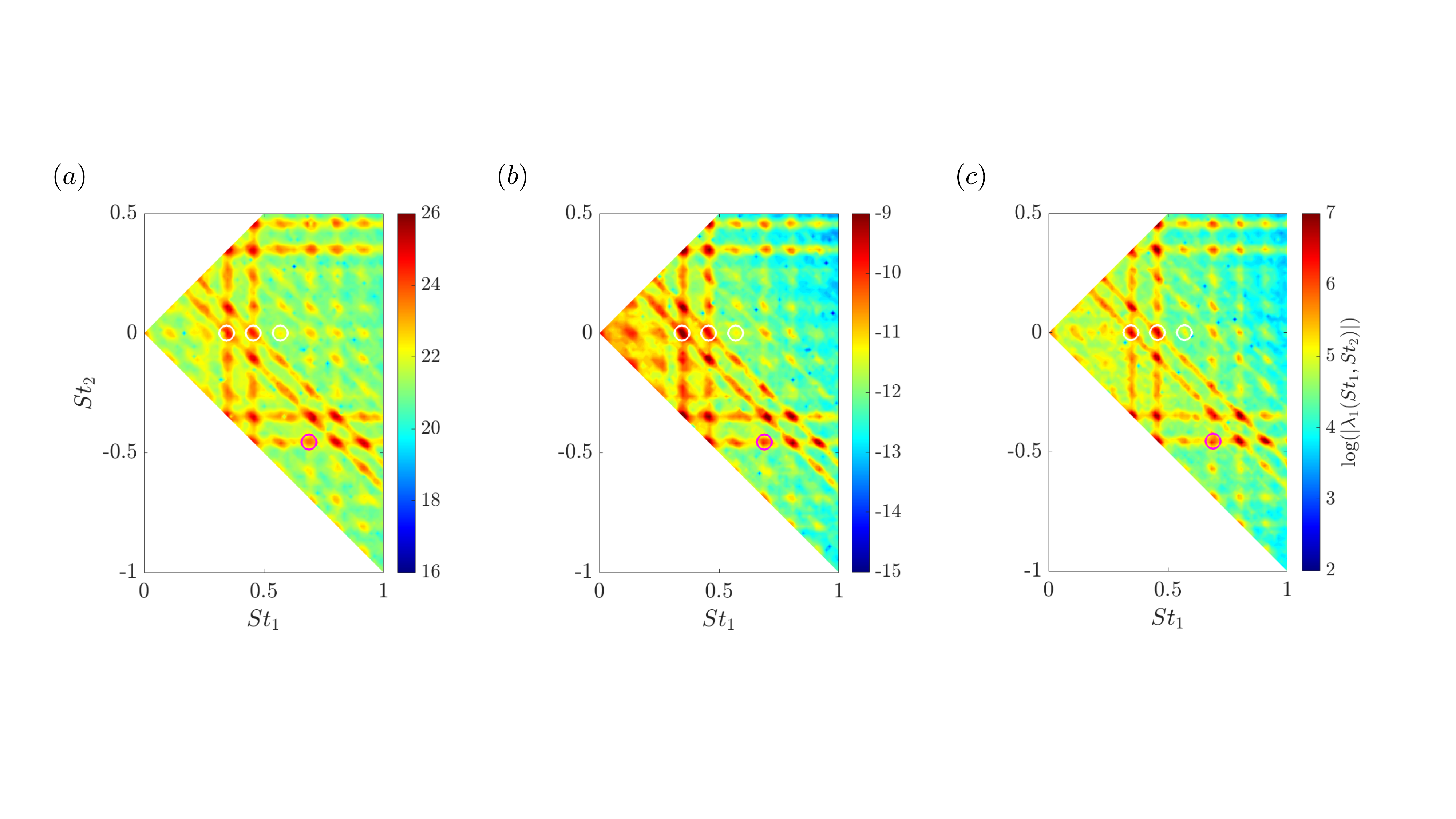}
\caption{Complex mode bispectrum based on (a) pressure; (b) density and (c) streamwise velocity. The white circles mark the frequencies of the three largest fundamental peaks, $St3=0.455$ (axisymmetric) $St2=0.345$ (helical) and $St4=0.57$ (helical). The pink circle highlights the production of a tone due to the triad $2St2-St3 \approx St1$. The generated tone can be observed in pressure spectra of case Jetideal3D, as shown in figure \ref{SPL_3dj}, and corresponds closely to the frequency $St1$.}
\label{bispod}
\end{figure}

Figure \ref{bispod} shows the complex bispectrum computed with pressure, density and streamwise velocity signals, separately for the Jetideal3D case. The three spectra display a similar grid pattern, produced by the self-interaction of the fundamental tones (originated from the resonant cycle) and their sum-and-difference combinations. Local maxima appear at the intersection of frequencies forming resonant triads. Three of the largest tones observed for this case are at $St2=0.345$ and $St3=0.455$ and $St4=0.57$ (see table \ref{tab:st} and figure \ref{SPL_3dj}); they correspond to fundamental tones generated by the resonant cycle, and are highlighted by white circles along the $St_2=0$ line. The pink circles highlight the triad formed by the first harmonic of $St2$ and $St3$, i.e. $2St2-St3 \approx St1$ (a little margin is expected due to spectral leakage). The third frequency generated in this triad corresponds closely to the frequency of the first major tone identified in the pressure spectrum. As discussed in \S \ref{sec:results}, this tone falls outside of the allowable frequency range for the existence of the upstream-travelling modes (see figure \ref{tones} for the $m=1$ mode) and therefore does not correspond to a fundamental frequency of the resonant cycle. The presence of a local maximum at the intersection of $2St3$ and $St2$ reveals that the $St1$ tone most likely issues from the nonlinear interaction of two fundamental modes, which explains the mismatch with the linear model. The mode bispectrum reveals a number of other sum and difference interactions between $St2$ and $St3$ that give rise to higher frequency tones observed in the pressure spectrum of figure \ref{SPL_3dj}.

\section*{Appendix B: Guided modes of a streamwise-averaged mean flow}
\label{sec:appendix2}

\begin{figure}[!hb]
\centering
\includegraphics[trim=0cm 0cm 0cm 0cm, clip=true,width=0.5\linewidth]{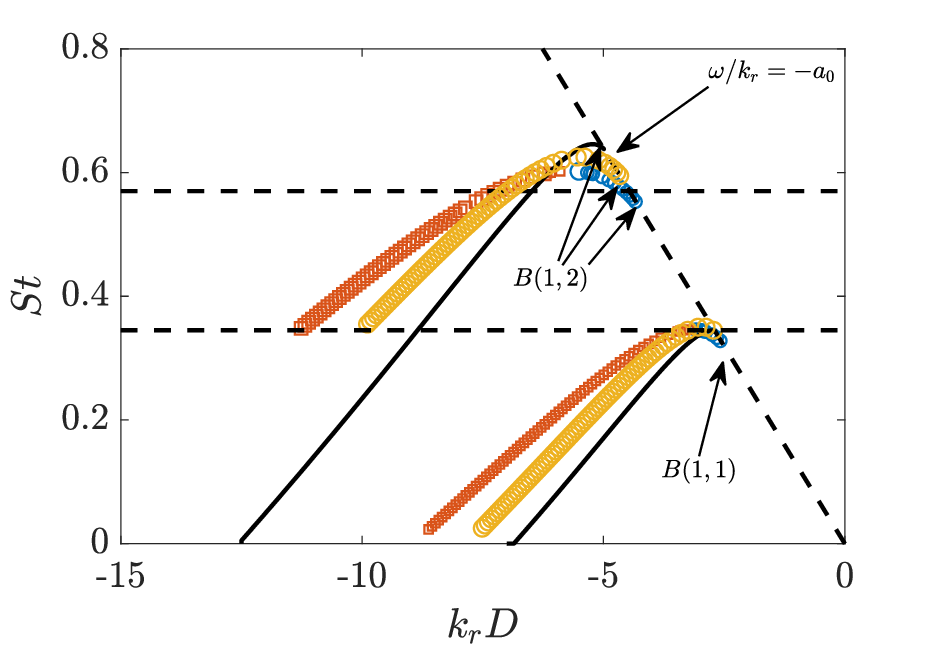}\caption{Dispersion relations of the neutral guided jet modes for azimuthal mode $m=1$. Solid black lines are vortex-sheet solutions. Blue and red circles correspond neutral with $R/\theta_R=5$ and the yellow points are the neutral waves obtained with the streamwise-averaged mean flow profile ($R/\theta_R=10$). Branch points of the different models are highlighted, respectively. The diagonal dashed line marks the sonic line. Horizontal dashed lines represent the frequency of the tones observed in the LES data.}
\label{disp_modif}
\end{figure}

In this section we present dispersion relations for the guided jet modes obtained with a spatially-averaged mean flow. First, we performed the streamwise average between $x/D=0$ and $x/D=3$ for the $L=3D$ case. The shear layer obtained with this averaged flow profile was then used to compute the $R/\theta$ parameter for the analytical mean flow, equation \ref{michalke_mean}, yielding a value of $R/\theta=10$, which is then used to perform a stability analysis. We recall that for the $m=1$ azimuthal mode, one of the observed tones is found to be well-predicted by the finite thickness model with $R/\theta=5$, but it is clearly off with respect to the vortex-sheet model (see figure \ref{tones}(a)). Figure \ref{disp_modif} shows the dispersion relation of the streamwise-averaged mean flow at $M_j=1.5$ and $m=1$ compared to that obtained with the vortex sheet model and with the finite-thickness model close to the plate location. The dispersion relation of the averaged mean flow falls between the other two, as expected. Furthermore, the frequency of the second tone (which is inconsistent with the vortex sheet) is quite close to the branch point associated with $R/\theta=10$. This suggests that this streamwise-averaged mean flow could be used as a \enquote{typical} profile representing a compromise between the downstream and upstream reflection points. This would allow avoiding the choice of one reflection point or the other to perform tone predictions.

Similar results can be obtained for other nozzle-to-plate distances and for the $m=0$ mode. Moreover, notice that for the other cases explored in the present work ($L=4D, 5D, 6D$), the value of $R/\theta_R$ associated with the spatially-averaged mean flow would be even closer to that obtained at the plate, as it varies very little for $x/D>2$ (see figures \ref{r_theta}(b), \ref{r_theta}(c) and \ref{r_theta}(b)).

\bibliographystyle{unsrtnat}
\bibliography{sample}

\end{document}